\documentclass[aps,prd,twocolumn,amsmath,amssymb,amsfonts,nofootinbib,superscriptaddress]{revtex4-1}
\usepackage{float}
\usepackage{comment}

\usepackage{graphicx}
\usepackage{dcolumn}
\usepackage{bm}
\usepackage{url}
\usepackage{subfigure}
\usepackage{float}
\usepackage{amssymb}
\usepackage{amsmath}
\usepackage{longtable}
\usepackage{rotating}
\usepackage{color}
\usepackage{xcolor}
\usepackage{epsfig}
\usepackage{epsf}
\usepackage{fancyhdr}
\usepackage{hyperref}

\def\be{\begin{equation}}
\def\bea{\begin{eqnarray}}
\def\eea{\end{eqnarray}}
\def\ee{\end{equation}}
\def\bi{\begin{itemize}}
\def\ei{\end{itemize}}

\def\bn{\begin{enumerate}}
\def\en{\end{enumerate}}

\def\be{\begin{equation}}
\def\ee{\end{equation}}
\def\bea{\begin{eqnarray}}
\def\eea{\end{eqnarray}}

\def\beq{\begin{equation}}
\def\eeq{\end{equation}}

\def\erfc{\mathrm{erfc}}
\begin{document}

\title{A method to search for long duration gravitational wave transients from isolated neutron stars using the generalized FrequencyHough}


\author{Andrew Miller}
\affiliation{INFN, Sezione di Roma, I-00185 Roma, Italy}
\affiliation{Universit\`a di Roma La Sapienza, I-00185 Roma, Italy}
\affiliation{University of Florida, Gainesville, FL 32611, USA}

\author{Pia Astone}
\affiliation{INFN, Sezione di Roma, I-00185 Roma, Italy}
\author{Sabrina D'Antonio}
\affiliation{INFN, Sezione di Roma Tor Vergata, I-00133 Roma, Italy}
\author{Sergio Frasca}
\author{Giuseppe Intini}
\affiliation{INFN, Sezione di Roma, I-00185 Roma, Italy}
\affiliation{Universit\`a di Roma La Sapienza, I-00185 Roma, Italy}
\author{Iuri La Rosa}
\affiliation{INFN, Sezione di Roma, I-00185 Roma, Italy}
\affiliation{Max Planck Institute for Gravitational Physics (Albert Einstein Institute), D-30167 Hannover, Germany}
\author{Paola Leaci}
\author{Simone Mastrogiovanni}
\affiliation{INFN, Sezione di Roma, I-00185 Roma, Italy}
\affiliation{Universit\`a di Roma La Sapienza, I-00185 Roma, Italy}
\author{Federico Muciaccia}
\affiliation{Universit\`a di Roma La Sapienza, I-00185 Roma, Italy}
\author{Cristiano Palomba}
\affiliation{INFN, Sezione di Roma, I-00185 Roma, Italy}
\author{Ornella J. Piccinni}
\affiliation{INFN, Sezione di Roma, I-00185 Roma, Italy}
\affiliation{Universit\`a di Roma La Sapienza, I-00185 Roma, Italy}
\author{Akshat Singhal}
\affiliation{INFN, Sezione di Roma, I-00185 Roma, Italy}
\author{Bernard F. Whiting}
\affiliation{University of Florida, Gainesville, FL 32611, USA}


\begin{abstract}
We describe a method to detect gravitational waves lasting $O(hours-days)$ emitted by young, isolated neutron stars, such as those that could form after a supernova or a binary neutron star merger, using advanced LIGO/Virgo data. The method is based on a generalization of the FrequencyHough (FH), a pipeline that performs hierarchical searches for continuous gravitational waves by mapping points in the time/frequency plane of the detector to lines in the frequency/spindown plane of the source. We show that signals whose spindowns are related to their frequencies by a power law can be transformed to coordinates where the behavior of these signals is always linear, and can therefore be searched for by the FH. We estimate the sensitivity of our search across different braking indices, and describe the portion of the parameter space we could explore in a search using varying fast Fourier Transform (FFT) lengths.
\end{abstract}


\maketitle

\section{\label{sec:level1}Introduction}

In the past two years, LIGO and Virgo have detected gravitational waves (GWs) from coalescing black holes and a binary neutron star system, which has opened the era of gravitational wave astronomy \cite{gw170817,gw150914,gw170814}. However, binaries are only one of many sources of gravitational waves. We focus here on signals emitted from a young, asymmetrically rotating neutron star, which could be a remnant of binary neutron star mergers or result from a supernova explosion \cite{yurem,aiparmspace}. There has been much work done on searching for signals of $O(s)$-bursts, binary inspirals, and for signals whose durations are quasi-infinite \cite{cwsearchO1,cwsearchO1two,fh_paper,powerflux,skyhough,fstatistic,keithreview}, but only a few pipelines have been developed to search for signals of intermediate duration, $O(hours-days)$, despite the fact that signals on these time scales could hold a lot of interesting information about neutron stars \cite{nsreview}. Additionally, as we progress further into the advanced detector era, we expect to see many more binary neutron star mergers \cite{mergerrates} and potentially supernovae, so we must have the tools ready to search for very young, isolated neutron stars.

Current pipelines that can search for intermediate duration signals include the Stochastic Transient Analysis Multi-detector Pipeline (STAMP) and the Viterbi algorithm. STAMP cross-correlates the data from two detectors to create time/frequency maps that are then analyzed to find clusters of excess power \cite{stamp}. The method employs two pattern recognition algorithms: seed-based and seedless clustering \cite{seedclustering,seedlessclustering} to cluster pixels that may be of the same origin together. It does not assume anything about the physics of the source, making it a powerful tool to search for gravitational waves immediately after a merger \cite{firstpostmergersearch}. The Viterbi algorithms are currently being developed to search for sources of intermediate duration, but have traditionally been used to search for continuous gravitational waves with spin-wandering using hidden Markov models \cite{viterbiaus}. Both pipelines are unmodeled searches, which means they are robust towards unknown signals. Another method is being developed called the Digital Power Law Tracker, a variation of the SkyHough, which is a modeled search that tracks the phase evolution of the signal using many templates \cite{dplt}. Additionally, adapting traditionally CW methods for long duration transient has been investigated in \cite{davidstat}, and a new detection statistic was derived.

Different physical mechanisms, described by the braking index. could be responsible for the spindown from a young neutron star. A braking index determines the frequency behavior of an expected signal from a neutron star as a function of time \cite{bi_of_pulsars}:

\begin{equation}
    n=\frac{f|\ddot{f}|}{\dot{f}^2}
\end{equation}
where $f$ is the rotation frequency of the neutron star, $\dot{f}$ is the spindown, and $\ddot{f}$ is the rate of change of the spindown.

The braking index has been measured in a few cases \cite{archibald,clark,lowbikarl,winds} but the physical mechanisms behind it are not well-understood. Some pulsars have been observed close to $n=3$, which is indicative that not just dipole electromagnetic radiation is causing the spindown of these pulsars \cite{archibald,clark}. Very recently, braking indices have been found to be around $n=2.5$ for millisecond pulsars born in GRB 130603B and GRB 140903A \cite{lowbikarl}. And some pulsars have braking indices as low as $n=1$ \cite{winds}. However, we expect to employ this method to search for neutron stars that are much younger than those whose braking indices have been measured.


This paper is organized as follows: in section \ref{sec:level2} we describe the physics of the general braking index model. In section \ref{sec:level3} we discuss the existing FH (FH) method, which is designed to search for continuous gravitational waves (CWs); in section \ref{sec:level4} we explain modifications to this method to search for pulsars whose time/frequency behavior follows a power law. We explain our procedure for doing coincidences and how we would follow up coincident candidates in section \ref{sec:level8}, and describe a set up of the parameter space for a real search in section \ref{sec:level5}. In section \ref{sec:level88} we discuss how we estimate the sensitivity of our search theoretically and using software injections for a range of braking indices, and in section \ref{sec:level11} we conclude by discussing improvements to the method and a possible application of machine learning.

\section{\label{sec:level2}Long duration transients}

We assume that radiation emitted by an isolated neutron star will follow a power law of the form \cite{bi_of_pulsars}:

\begin{equation}
    \dot{f}=-k f^n
    \label{diff_powlaws}
\end{equation}
where $k$ is a proportionality constant that contains the physics of the emitted radiation, defined to be positive. This radiation is modelled as coming from the rotational energy of the neutron star; hence as the energy is emitted, the neutron star spins down.

Different braking indices correspond to different physical mechanisms. $n=1$ refers to wind braking, which is an electromagnetically powered flux of relativistic particles away from the neutron star. The mechanism behind pulsar winds is not well understood, but it is believed that the extremely fast rotation of a magnetized neutron star causes a large electric field that rips particles from the surface of the neutron star and accelerates them to relativistic speeds. The case $n=3$ refers to magnetic braking: in the simplest models, neutron stars are treated as having a time-varying magnetic dipole moment, which is used to then find the power emitted as a function of time \cite{nstextbook}. The case $n=5$ is braking due to gravitational wave emission from a ``mountain'' or ``crater'' on the surface of an otherwise spherically symmetric neutron star \cite{gwquad}. which causes a nonaxisymmetric ($I_x\neq I_y$) neutron star rotating about its other principle axis ($I_z$). The case $n=7$ refers to gravitational waves emitted by r-modes, modes that result from small velocity and density perturbations of the neutron star fluid that cause a time-varying moment of inertia. The restoring force for r-modes is the Coriolis force, hence these modes only occur on rotating bodies and are Rossby waves in the earth's atmosphere \cite{owen1998}. However, it is possible that a combination of mechanisms occurs, and therefore the braking indices will not be equal to 1, 3, 5, and 7.

Integrating Eq. \eqref{diff_powlaws}, we find the frequency evolution as a function of time:

\begin{equation}
f(t)=\frac{f_0}{\left(1+k (n-1)f_0^{n-1}(t-t_0)\right)^{\frac{1}{n-1}}}
\label{powlaws}
\end{equation}
where $f_0$ is the rotational frequency at time $t_0$, and $t_0$ is also the time when our analysis begins. For all braking indices except $n=7$ r-modes (r-modes), we model the amplitude change as a function of time and frequency as \cite{cwamp}: 

\begin{equation}
    h(t)=\frac{4\pi^2 G}{c^4}\frac{\epsilon I_{z}}{d}{f(t)}^2
\end{equation}
where $\epsilon$ is the ellipticity of the neutron star, $I_{zz}$ is the moment of inertia with respect to the star's rotation axis, $d$ is the distance to the source, $G$ is Newton's gravitational constant, and $c$ is the speed of light.

For r-modes, we use the following relation \cite{rightrmode}:



\begin{equation}
    h(t)=\sqrt{\frac{2^{15} \pi^7}{5}}\frac{G J M R^3}{d}\alpha f(t)^3
\end{equation}
where $\alpha$ is the saturation amplitude, the amplitude at which  non-linear effects stop the growth of the r-mode, $J=1.635 \times 10^{−2}$ is a dimensionless angular momentum for polytropic models, and $M$ and $R$ are the mass and radius of the neutron star.

We assume $k$ is independent of time, which is reasonable in the case of a fixed braking index or one that very slowly varies in time:

\begin{equation}
    k=\frac{|\dot{f}_0|}{f_0^n}
\end{equation}
where $\dot{f}_0$ is the initial spindown of the source.

\section{\label{sec:level3}Original FrequencyHough Transform}

The hierarchical pipeline described in \cite{fh_paper} and used in \cite{fh_search} performs searches for continuous gravitational waves from asymmetrically rotating neutron stars. The method employs a particularly efficient implementation of the standard Hough transform \cite{houghpatent}. In the first step, the raw strain data are cleaned and stored in so-called short FFT databases (SFDBs) \cite{sfdb_paper}. For each interlaced FFT in the SFDB, the average noise is estimated based on an auto-regressive mean that tracks the variation in the noise without being sensitive to peaks in the spectrum. This is important in the next step, when the peakmap is constructed. The peakmap is a collection of time/frequency points whose amplitudes -the signal power in the spectrum- are above a threshold and a local maximum. An example of a peakmap is shown in the left panel of figure \ref{pm}. The (Doppler-corrected) peakmap is the input to the Hough transform.

The FH transforms points in the time/frequency plane of the detector to lines in the frequency/spindown plane of the source. The CW signal is modeled as a monochromatic signal with a small spindown:

\begin{equation}
f = f_0+\dot{f}(t-t_0)+\frac{\ddot{f}}{2}(t-t_0)^2+\ldots
\label{Taylor}
\end{equation}
$t$ is the time at which we observe the Doppler corrected frequency $f$ on earth; $\ddot{f}$ is the second order spindown of the source. 

The decrease in gravitational wave frequency originates from a small deformation on the star's surface, caused by its inner magnetic field \cite{innerbfield}. Modifications to this deformation can include changes in the magnetic field or starquakes.

For each position in the sky, each peak ($t-t_0$,$f$) in the peakmap will be transformed into a line in the frequency/spin down plane ($f_0$,$\dot{f}$):

\begin{equation}
\dot{f}= -\dfrac{f_0}{(t-t_0)}+\dfrac{f}{(t-t_0)}
\label{fdotslope}
\end{equation}
Candidates present in the peakmap are formed by a superposition of several lines in this new space. The second order spindown correction in \eqref{Taylor} is neglected since it is assumed to be very small, and even if this parameter is considered, we can correct for it at the level of the peakmap by simply shifting the frequencies at which we receive peaks.


\section{\label{sec:level4}Generalized FrequencyHough Transform}

In order to search for signals with highly varying frequencies, we cannot apply the traditional FH, since the frequency/time evolution of the signal is not linear (see Eq. \eqref{powlaws}). Therefore, we must transform the nonlinear equations into a coordinate system in which the behavior of the signal is linear. The transformation is:

\begin{equation}
x=\frac{1}{f^{n-1}}; x_0=\frac{1}{f_0^{n-1}}
\label{transform}
\end{equation}
Under this transformation, Eq. \eqref{powlaws} becomes the equation of a line:

\begin{equation}
x=x_0+(n-1)k(t-t_0)
\end{equation}
Points in the ($t-t_0$,$x$) plane are mapped to lines in the ($x_0$,$k$) plane: 

\begin{equation}
    k=-\frac{x_0}{(n-1)(t-t_0)}+\frac{x}{(n-1)(t-t_0)}
\end{equation}
We now have analogous equations to equations \ref{Taylor} and \ref{fdotslope}.

\begin{figure*}[ht!]
\centering
\begin{minipage}[b]{.4\textwidth}
\includegraphics[width=80mm]{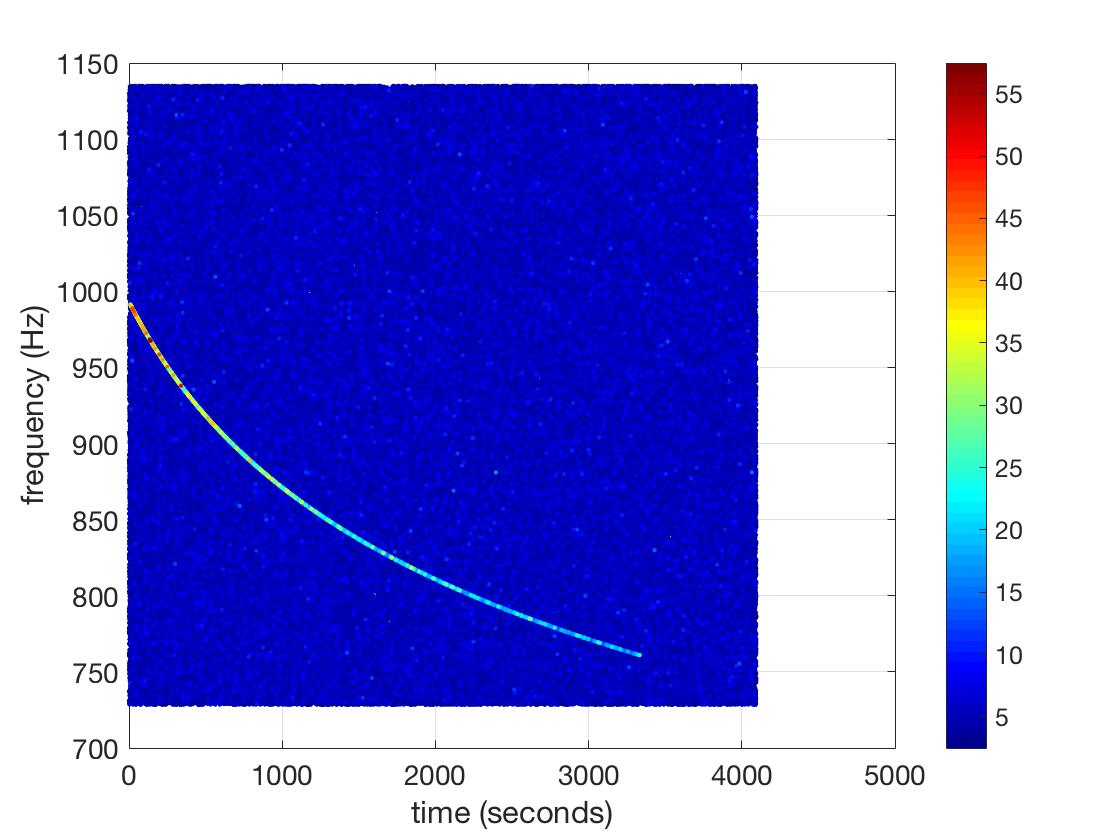}
\end{minipage}\qquad
\begin{minipage}[b]{.4\textwidth}
\includegraphics[width=80mm]{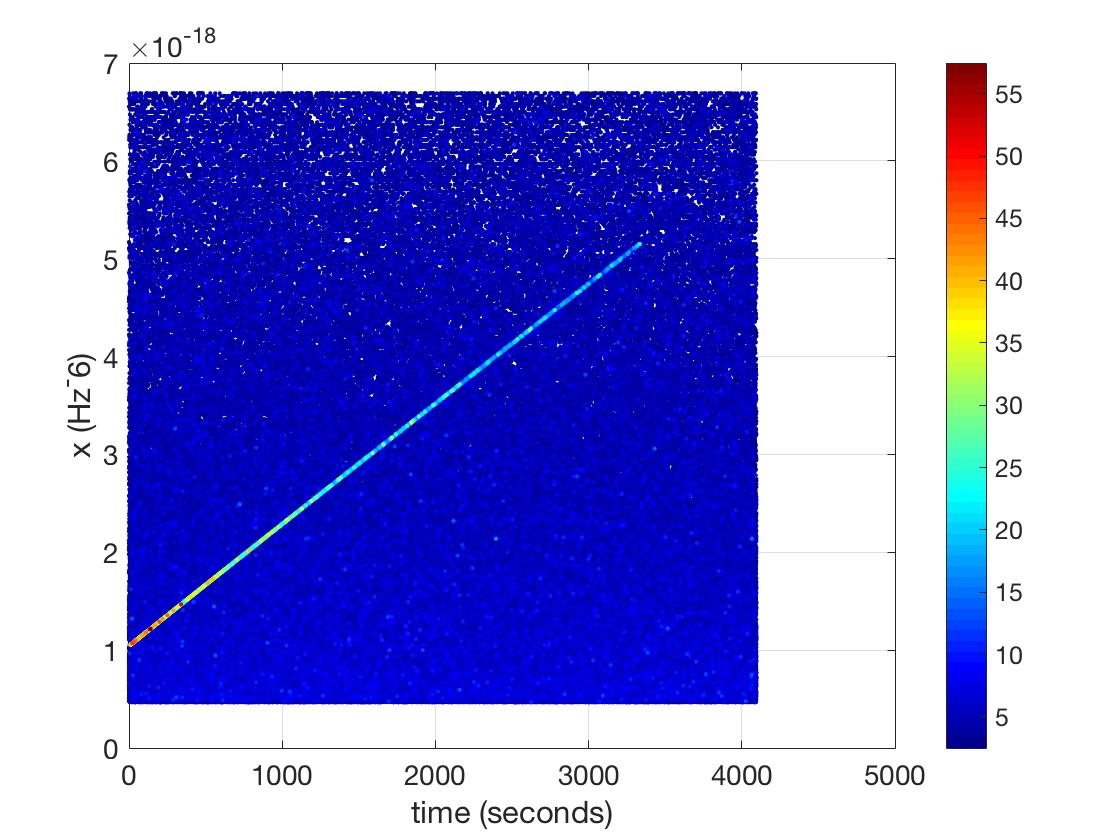}
\end{minipage}
\caption{Left: peakmap of injected r-mode ($n=7$) signal in white noise with $h_0=1.5\times 10^{-22}$, $f_0=993.25$ Hz, $\dot{f_0}=-1.99\times 10^{-1}$ Hz/s. Right: peakmap after transformation given by Eq. \eqref{transform} with parameters: $x_0=4.706\times 10^{-19}$ Hz$^{-6}$ and $k=9.458\times 10^{-23}$ Hz$^{-5}$. Signals in the time/frequency domain are transformed into lines in the time/x plane if the transformation (the braking index) is correct. The peakmap was constructed with an FFT length $T_{FFT}=2$ s.} 
\label{pm}
\end{figure*}

The right panel of figure \ref{pm} shows the transformed peakmap before it is fed to the FH transform. Notice that the peaks are denser near the low values of $x$ (which correspond to higher frequencies). This effect is also seen in the FH map; it occurs because in this new space, the noise is not uniform. The change of coordinates takes points that are equally spaced in frequency and concentrates them at higher frequencies (lower $x$ values) and spreads them out at lower frequencies (high $x$ values). So in the FH, more lines occur at lower $x$ values. Therefore, we construct peakmaps in a larger frequency band than we plan to actually analyze (20-25$\%$ larger). For example, if we wish to analyze the band [200 300] Hz, we create a peakmap spanning [200, 350] Hz. But when we select candidates after performing the generalized FH, we do not include the candidates in $x_0/k$ space that correspond to the extra 50 Hz. In a real search, the frequency bands we analyze are interlaced--the bands have some amount of overlap-- so that we can still cover all the frequencies. This is by far the simplest way to deal with the non-linearity of the noise in this new space at the cost of a moderate increase in computational cost.

\begin{figure}[ht!]
    \centering
    \includegraphics[width=90mm]{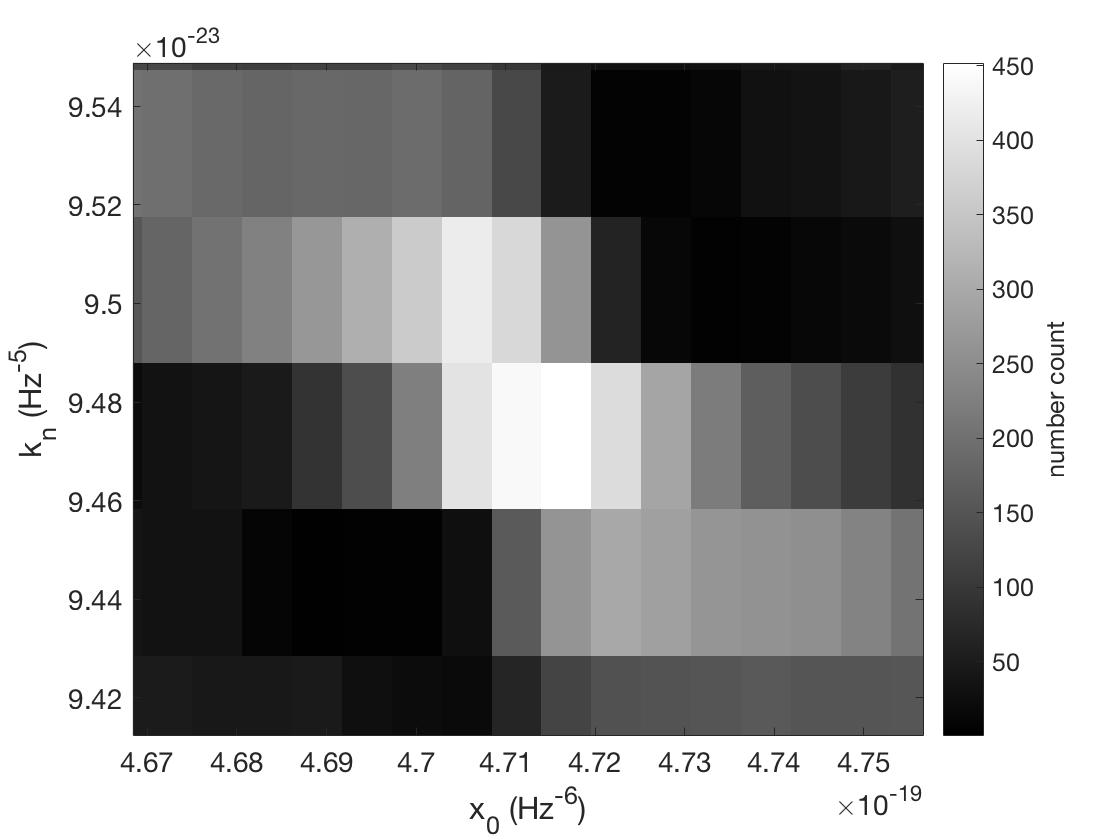}
    \caption{Hough map clearly showing the injected signal recovered, with the parameters given in the caption of figure \ref{pm}. Lower values of $x_0$ correspond to higher frequencies, and vice-versa. The color corresponds to the number of lines that hit each $x_0$/$k$ bin. The reference time $t_0$ is taken to be at the beginning of the time segment analyzed.}
    \label{hm}
\end{figure}

The output of the generalized FH transform is a histogram in $x_0/k$ space, where the highest number count refers to the most likely parameters ($x_0$ and $k$) of the signal contained in the peakmap. See figure \ref{hm}.

Several Generalized FHs will be performed, each one at a different $n$ and covering a range of different $k$. In each of the FH maps, we have to select significant candidates. Each candidate consists of a value for $x_0$, $k$, $n$, longitude and latitude. From $x_0$ and $k$, we can find $f_0$ and $\dot{f_0}$ with the inverse transformation of Eq. \eqref{transform}. We use a slight modification of the procedure to select candidates that is described in \cite{fh_paper}. We select candidates in each ``square'' of the Hough map, whose size (in bins) is determined by the number of candidates we wish to select in each map. Each axis of the Hough map is divided by the square root of the number of candidates that we want, creating separate regions in the Hough map from which to choose candidates. We employ this way of selecting candidates because we want to be cover the parameter space as uniformly as possible, and so that we are not blinded by major disturbances, nor the non-uniformity of the noise. As an example, we make 25 squares (meaning that the $x_0$ and $k$ axes are each split into 5 chunks) and select 1 candidate per square across the Hough plane with the highest number count, with the possibility of selecting another candidate in each square if it is sufficiently far away in number of $k$ bins $(\sim 3$ bins). Selecting a second candidate in each square (or strip) is done to not lose a potential candidate and is already used in the traditional method. The only difference is that for CW hierarchical searches with the FH, another candidate is selected if it is sufficiently far away in frequency, not spindown. 

In both the original and generalized FHs, we evaluate a detection statistic called the critical ratio to determine if a candidate is significant:

\begin{equation}
    CR=\frac{y-\mu}{\sigma}
\end{equation}

where $y$ is the number count in the Hough plane in a particular bin (a potential candidate), $\mu$ and $\sigma$ are the average and standard deviation of the number counts in the map due to noise, 

Once the analysis has been performed on all detectors and a set of candidates has been produced form each one, we then perform coincidences, which simply find if candidates selected in each detector are close to each other in the parameter space. We define ``close'' in the parameter space as being within a certain number of $x_0$ and $k$ bins. For the candidates in each detector, the distance between them in the parameter space is calculated in the following way, as was done in \cite{fh_paper}:

\begin{equation}
d=\sqrt{\left(\frac{x_{0,2}-x_{0,1}}{\delta x_0}\right)^2+\left(\frac{k_{2}-k_{1}}{\delta k}\right)^2}
\end{equation}
where $\delta x_0$ is the size of the bins in $x_0$, and $\delta k$ is the varying size of the bins in $k$ (see \ref{sec:level5}). Note that in general we should also include error in the sky location, but for now we are only considering a search where the position of the source is fixed. 

In a real search, if the candidates' parameters are within a distance of 3 bins of each other, we consider the candidates to be in coincidence. Three bins is the number used in the FH hierarchical searches and represents a compromise between achieving high detection efficiency while keeping the false alarm rate as low as possible. Additionally, we remove known noise lines from the data, and apply persistancy vetoes as described in \cite{fh_paper}.


The generalized FH should be used for searches in which a potential source's sky position is known. For example, the position could be constrained by electromagnetic observations after a supernova or a binary neutron star merger. A blind all-sky search would add a huge computational burden to the analysis, since already the parameter space of braking indices and $k$ values explored is quite large.

\section{\label{sec:level8}Follow-up}

The coincidences will eliminate most candidates, so those that remain are potential gravitational wave signals. Since the generalized FH transform gives us an estimate for $f_0$, $\dot{f}_0$, $n$ and $t_0$, we can go back to the original strain $h(t)$ data and correct for the phase evolution of the signal. We make a heterodyne correction $h’(t)=h(t)e^{j\phi_{sd}(t)}$, where the phase $\phi_{sd}$ is \cite{bsd}:

\begin{equation}
    \phi_{sd}(t)=2\pi \cdot\int^{T_{obs}}_{t_0}\frac{f_0}{\left(1+k (n-1)f_0^{n-1}(t’-t_0)\right)^{\frac{1}{n-1}}} dt’+\phi_{t_0}
    \label{phcor}
\end{equation}
In a perfect situation, this correction would cause the signal to remain at a single frequency (despite the size of the bin) for its duration in the time/frequency peakmap, meaning that longer FFTs could be used, as in CW searches. In practice, we know that the correction will not be perfect because the discretization in the parameter space causes us to lose precision on $f_0$ , so we will have a relatively constant frequency line in the peakmap that will spin up or spin down from an $f_0$. 

\begin{figure}[ht!]
    \centering
    \includegraphics[width=90mm]{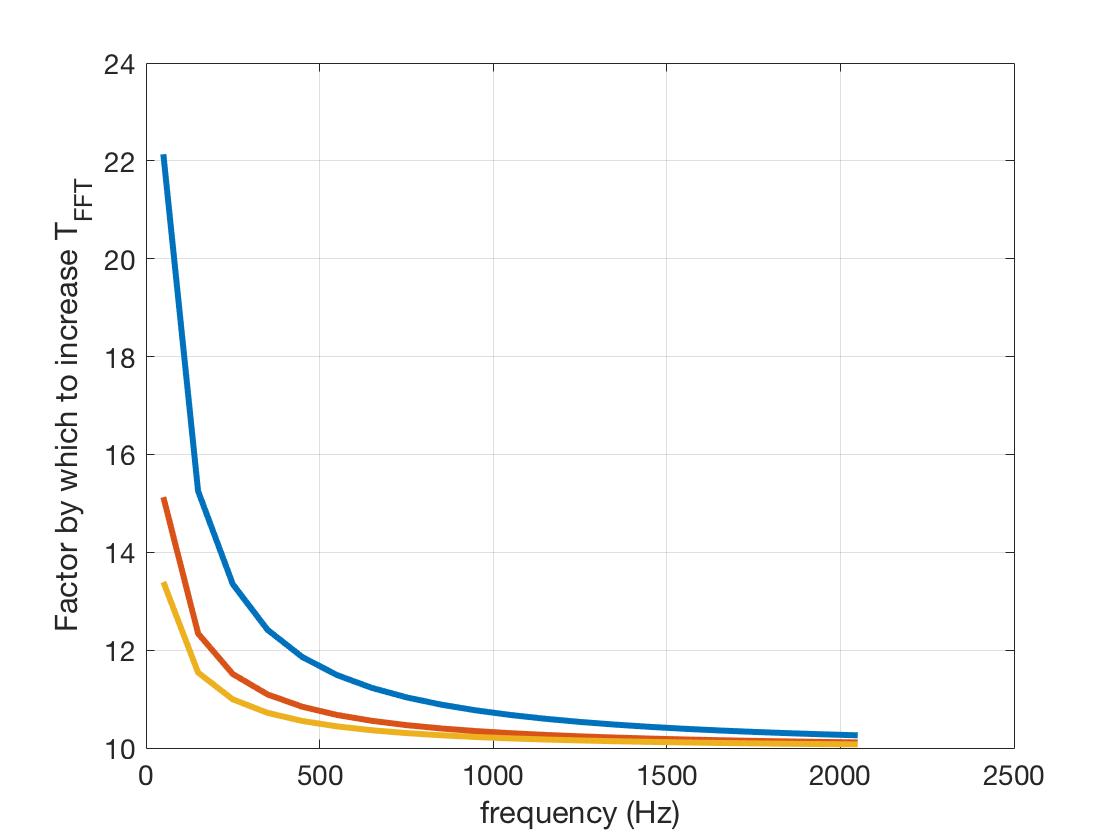}
    \caption{We chose the original $T_{FFT,0}=2$ s, $\dot{f_0}=0.25$ Hz/s. We show on the y-axis the factor by which we can increase $T_{FFT}$ during the follow-up as a function of frequency such that a signal would still be confined to one new frequency bin if we use an overresolution factor of 10 in both $x_0$ and $k$. Blue, red, yellow $\rightarrow n=3,5,7$ (also top, middle, bottom)}
    \label{enhance_tfft}
\end{figure}

How much we can increase the FFT length $T_{FFT}$ is determined by how much of an increase in computation time we can handle, as well as the size of residual spindowns. When we select candidates and do coincidences, we allow a coincidence window of 3 bins. If the signal is actually within three bins, we must construct a grid around each recovered parameter such that the signal's true parameters will be further isolated to one of the (smaller) bins in the follow-up. As an example, if we increase the resolution in both $x_0$ and $k$ by at least a factor of 10, then we can increase $T_{FFT}$ by a factor of 10. See figure \ref{enhance_tfft}.

\begin{figure*}[ht!]
\centering
\begin{minipage}[b]{.4\textwidth}
\includegraphics[width=80mm]{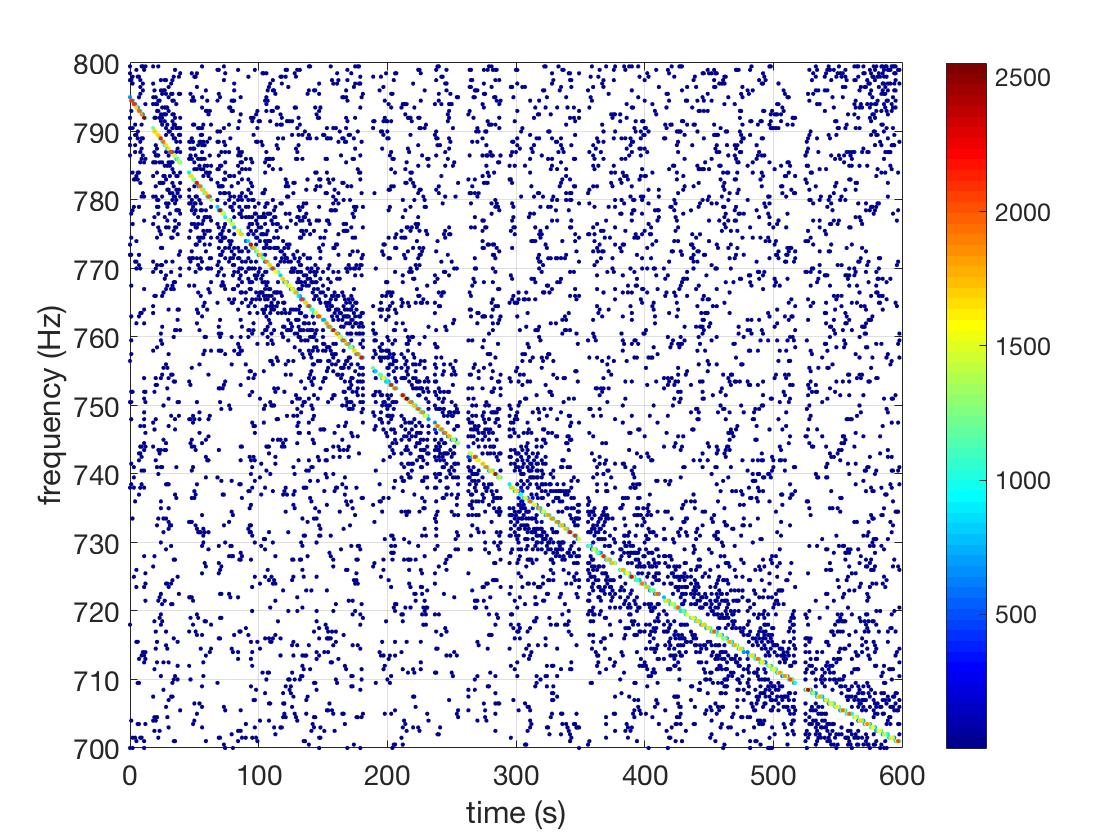}
\end{minipage}\qquad
\begin{minipage}[b]{.4\textwidth}
\includegraphics[width=80mm]{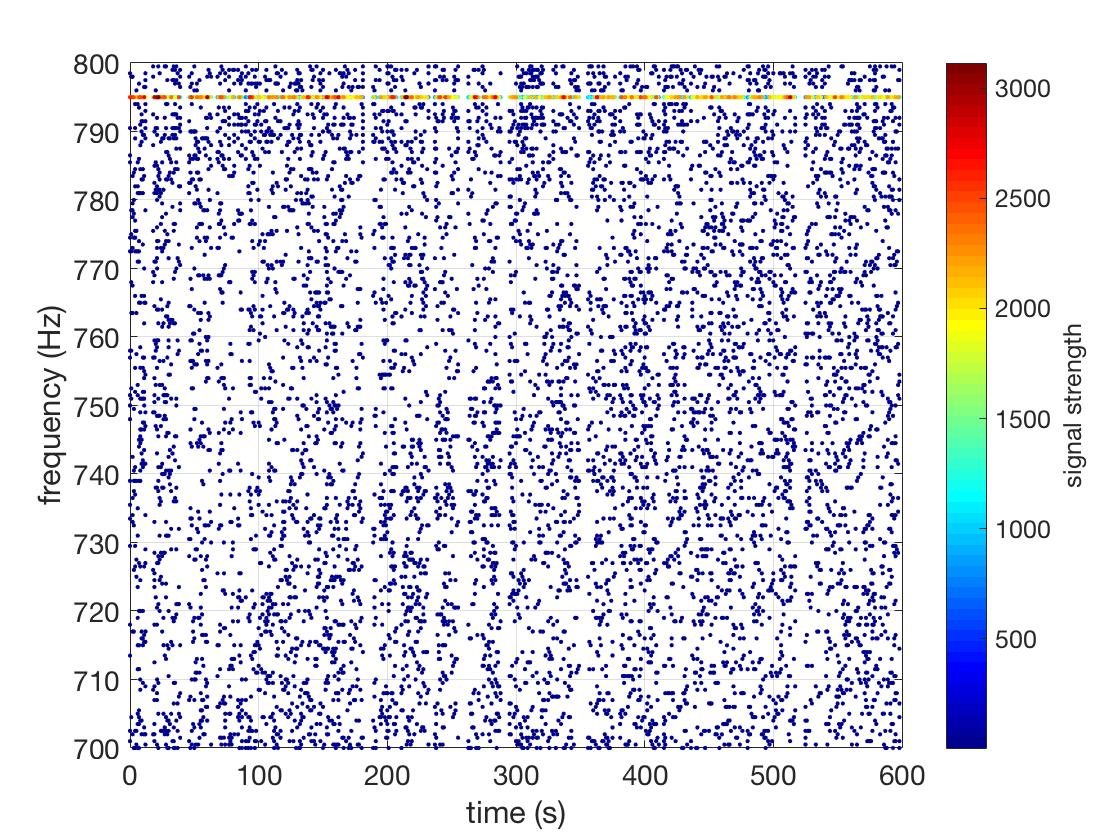}
\end{minipage}
\caption{Left: peakmap of injected r-mode ($n=7$) signal in white noise with $h_0=1\times 10^{-22}$, $f_0=795$ Hz, $\dot{f_0}=-2.5\times 10^{-1}$ Hz/s. Right: peakmap after phase evolution correction; a monochromatic signal is seen at $f_0=795$ Hz. The peakmap on the left was constructed with $T_{FFT}=2$ s; on the right, as is the case of the follow-up, we used $T_{FFT}=4$ s, twice as large as in the original analysis. The signal was injected and its phase corrected using the BSD framework.} 
\label{phasecorr}
\end{figure*}

In practice, we will perform the follow-up using Band-Sampled Data (BSDs), which are flexible data structures developed to efficiently handle data \cite{bsd}. In this framework, it is very easy to correct for the phase evolution of a potential signal and to change the $T_{FFT}$. Imagine that we run a search using the generalized FH and recover a candidate at $n=7$ (r-mode) with $f_0=795$ Hz and $\dot{f_0}=-0.25$ Hz/s using an initial $T_{FFT,0}=2$ s. The signal would appear in the peakmap (before the Generalized FH is applied) as in the left-hand panel of figure \ref{phasecorr}, which is an injection done within the BSD framework. Knowing these parameters, we can then correct the signal's phase evolution and use longer FFTs, since the signal's residual spindown will be lower. In the right panel, by applying Eq. \eqref{phcor}, we obtain a monochromatic signal, where we have used $T_{FFT}=2T_{FFT,0}$. After correcting for the phase, we then apply the original FH \cite{fh_paper} because any residual spindown remaining (when the phase correction is not perfect) could be small enough such that the time/frequency behavior is linear. We have done tests by correcting for the wrong frequency, spindown and/or braking index, and have found that if the parameters are off by around one bin at the most, we can still recover the signal after coincidences. 

After performing the original FH, we select candidates in this Hough map, where a candidate is defined by having a frequency, spindown and sky location, and do coincidences between the two detectors. If there is no coincidence, the candidate is vetoed; otherwise, we use the candidates' recovered parameters to correct the peakmap that was fed to the original FH (the right panel of figure \ref{phasecorr}). This correction requires a simple shifting of bins, which is done for the Doppler correction. After this second correction, we project this peakmap onto the frequency axis and calculate a new critical ratio:

\begin{equation}
    CR=\frac{y-\mu}{\sigma}
\end{equation}
where $x$ is now the number of peaks in the peakmap at the estimated signal initial frequency, $\mu$ is the mean of the number of peaks in the peakmap due to noise, and $\sigma$ is the standard deviation of the noise.

\begin{figure*}[ht!]
\centering
\begin{minipage}[b]{.4\textwidth}
\includegraphics[width=80mm]{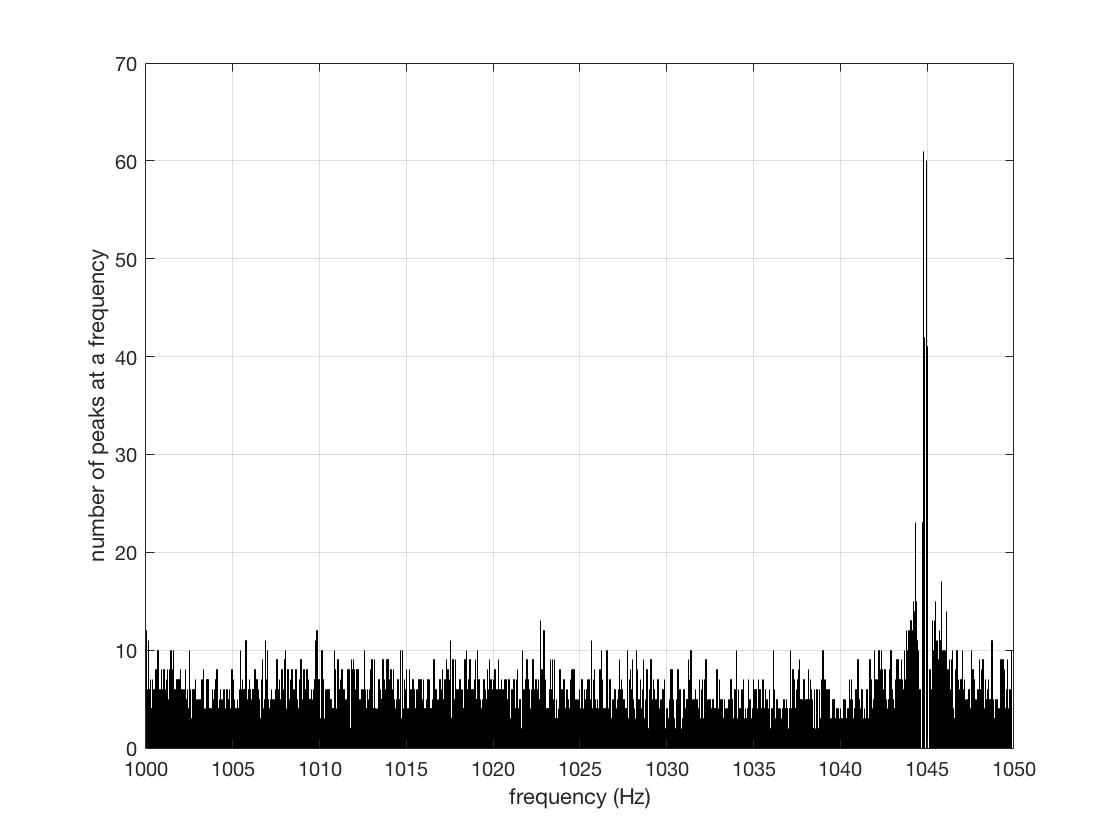}
\end{minipage}\qquad
\begin{minipage}[b]{.4\textwidth}
\includegraphics[width=80mm]{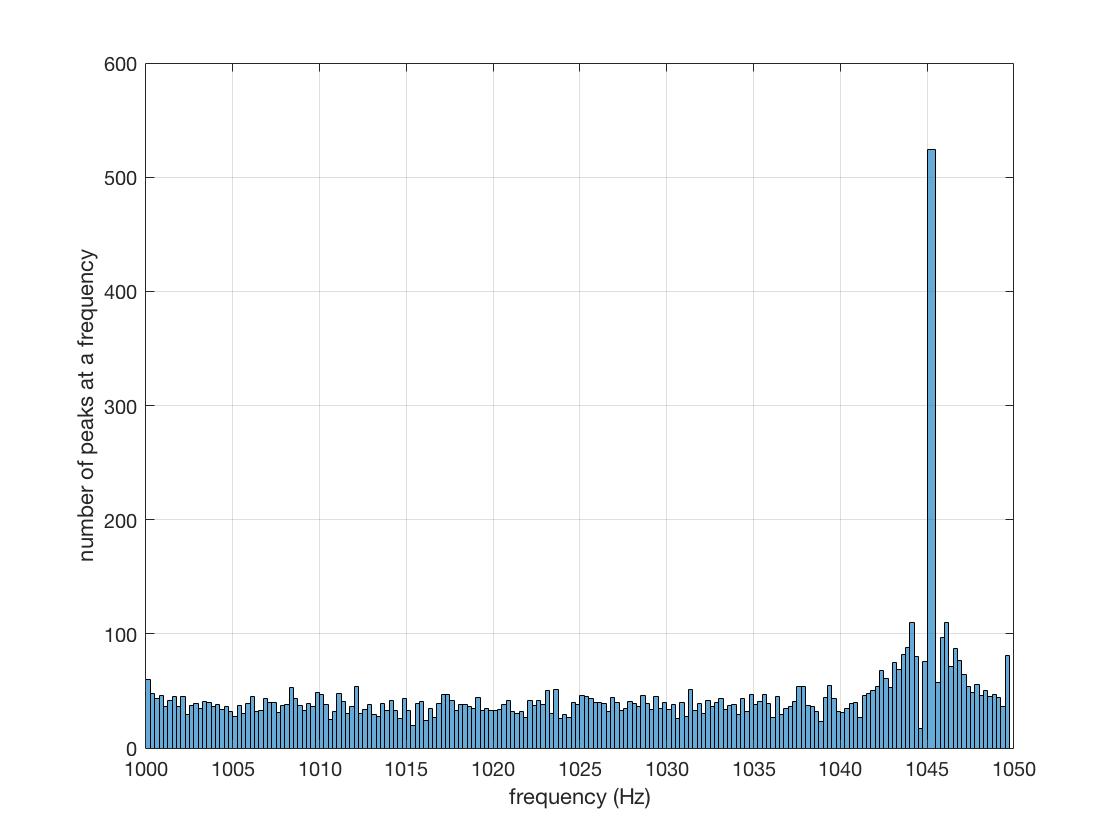}
\end{minipage}
\caption{A signal was simulated with $f_0=1045$ Hz,$\dot{f_0}=-0.03125$ Hz/s, $n=7$, and $h_0=8\times 10^{-23}$ in white noise. Left: we performed the Generalized FH, obtained candidates, and corrected the phase evolution in original peakmap $T_{FFT,0}=4 $ s, then projected that peakmap onto the frequency axis. Right: we then corrected for the phase evolution of this signal while increasing the $T_{FFT}$ by a factor of 2, ran the original FH, and used the best candidate's parameters to shift the bins in the peakmap as we do for the Doppler shift. In one detector, the critical ratio is increased by about 20\%; when using both detectors, the weighted average critical ratio is increased by about 22\%, which roughly equals $T_{FFT}^{1/4}$} 
\label{proj_pms}
\end{figure*}

We now need to compare this critical ratio to the critical ratio of the candidate before the follow-up. However, in a real search, the original peakmap spans $O(100)$ Hz, and the peakmap created in the follow-up only spans 10 Hz, since we have corrected for the signal's frequency evolution. Directly comparing these peakmaps is not fair because the average noise is different, and the original peakmap contains a signal with rapidly varying frequency, so its power is spread across many bins. Therefore, we must recompute the original peakmap (with $T_{FFT,0}$) in the same 10 Hz band, and correct for the phase evolution, so that the original signal would be confined to one frequency bin. We can then project this peakmap onto the frequency axis and compare this projection to that done with $2T_{FFT,0}$. The left panel of figure \ref{proj_pms} shows, for one detector, the projection of the peakmap created with $T_{FFT,0}$; the right panel shows the projection of the peakmap made with $T_{FFT}=2T_{FFT,0}$, that has been created with the follow-up procedure.


This process is done for each detector individually. We then weight the critical ratio by each detector's sensitivity around the candidate frequency, and compute an average critical ratio. If this weighted average critical ratio is greater than that computed on the original $T_{FFT,0}$ peakmaps by a factor of $T_{FFT}^{1/4}$, then we have a candidate that requires more analysis; otherwise, the candidate is vetoed.


\section{\label{sec:level5}Analysis of the parameter space}

\subsection{Grid on braking index $n$}

The search is model-dependent, and yet the power law models for neutron stars are quite uncertain. Therefore we search across different braking indices, which can correspond to either a combination of emission mechanisms- electromagnetic and gravitational wave emission from a neutron star -- or a braking index that is very slowly varying in time. We construct a grid of different braking indices between $n=2.5$ and $n=7$ with a step that is calculated so that the frequency variation stepping from $n_1=n$ to $n_2=n+dn$ for the signal duration is confined to one frequency bin:

\begin{eqnarray}
f_1(t)&=&\frac{f_0}{\left(1+k (n_1-1)f_0^{n_1-1}(t-t_0)\right)^{\frac{1}{n_1-1}}} \\
f_2(t)&=&\frac{f_0}{\left(1+k (n_2-1)f_0^{n_2-1}(t-t_0)\right)^{\frac{1}{n_2-1}}} \\
\Delta f &=& f_2(t)-f_1(t) \leq \delta f=\frac{1}{T_{FFT}}
\label{dn_powlaws}
\end{eqnarray}
We empirically find the value of $dn$ at each braking index $n$, $k$ and $f_0$ such that for the duration of the analysis $T_{obs}$, the frequency variation remains within one frequency bin. $\delta f$ is fixed by our choice of $T_{FFT}$, which depends on the maximum spindown we wish to analyze. Therefore, the grid depends on the FFT length, the braking index, the spindown range and frequency range.

\begin{figure}[ht!]
\centering
\includegraphics[width=80mm]{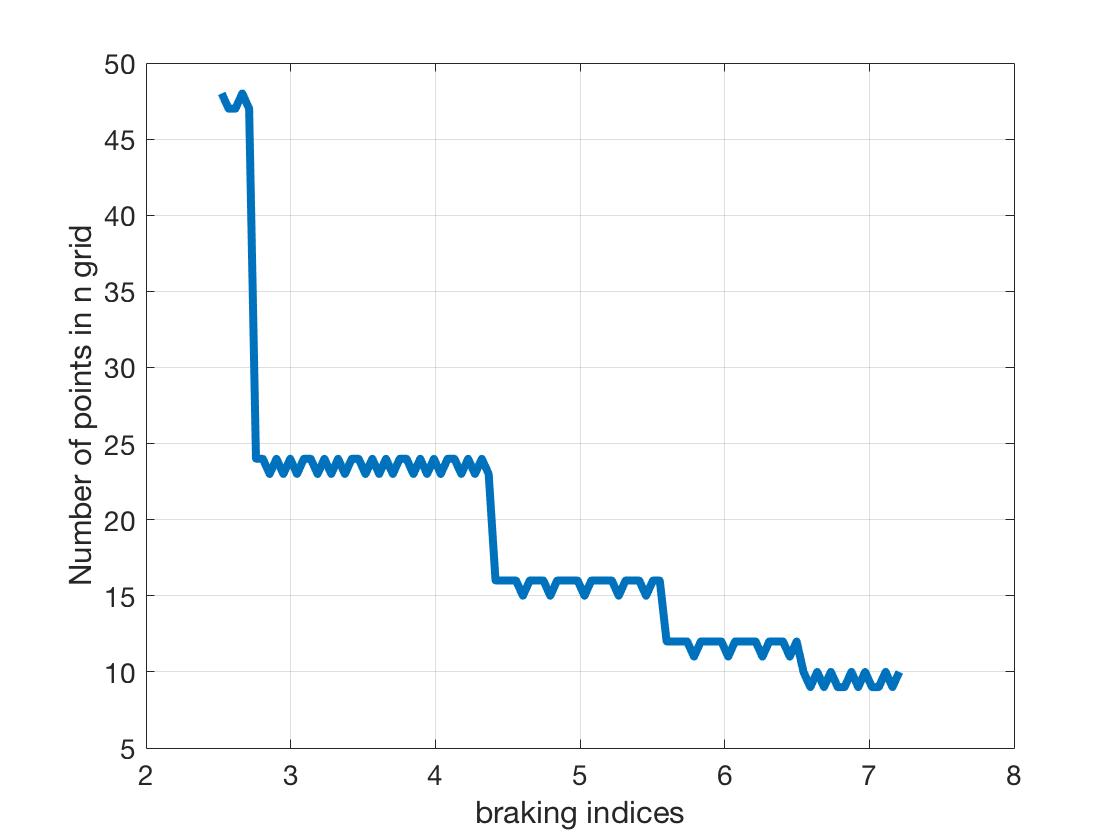}
\caption{The plot shows the number of points in the grid on $n$ as a function of the braking index. Parameters used in this plot: $T_{FFT}=2$ s, $f_{0,max}=2000$ Hz, $\dot{f}_{0,max}=1/T_{FFT}^2$ $n=[2.5,7]$, histogrammed into 100 $n$ bins of size $\delta n$ $\sim 4\times 10^{-2}$.}
\label{npts}

\end{figure}

We provide here some plots describing the behavior of the grid on $n$. In figure \ref{npts}, we show how the number of points changes in the grid as a function of the braking index. At higher braking indices, the magnitude of the spindown decreases quicker than at lower braking indices, meaning that fewer points in the grid $n$ are required to cover the parameter space. 

\begin{figure}[ht!]
    \centering
    \includegraphics[width=90mm]{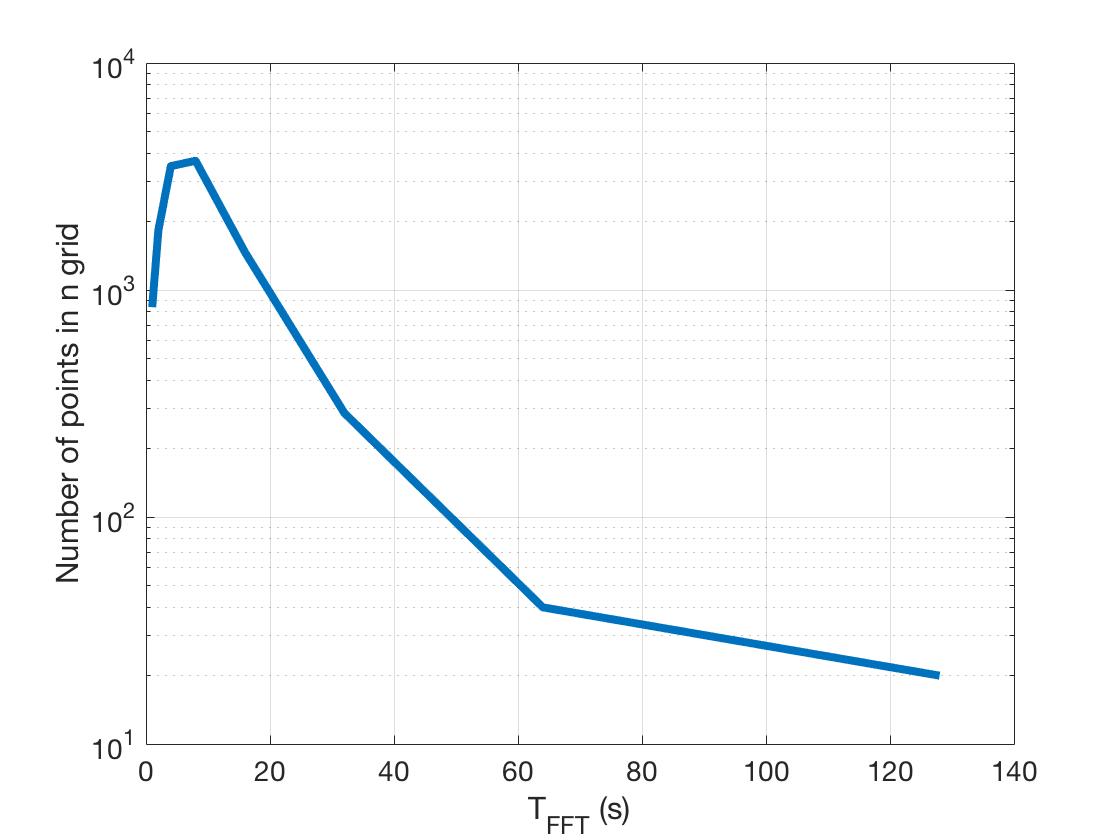}
    \caption{This plot shows the number of points in the grid on $n$ as a function of $T_{FFT}$ for a fixed observation time $T_{obs}=10^5$ s and maximum frequency $f_{0,max}=2000$ Hz. Note that for longer $T_{FFT}$, the spindowns we are analyzing are smaller. We see a sharp decay in the number of points in the grid because the frequency is not varying so strongly, and therefore Eq. \eqref{powlaws} reduces to Eq. \eqref{Taylor}.}
    \label{n_vs_tfft}
\end{figure}

Additionally, the number of points in this grid changes as a function of $T_{FFT}$ for two reasons: (1) the frequency bin size becomes smaller with increasing $T_{FFT}$ and (2) the frequency dependence as a function of time becomes much weaker as the spindowns we analyze decrease (since $\dot{f}=1/T_{FFT}^2$ and $T_{FFT}$ is increasing), see figure \ref{n_vs_tfft}. 

\subsection{Grid on proportionality constant $k$}

We must also construct a grid in $k$, since the Hough map is constructed across all times for a fixed $k$. An expression for the step can be derived analytically if we use $log_{10}$ of Eq. \eqref{diff_powlaws}:

\begin{eqnarray}
    \dot{f}&=&-k f^n
    \label{eqn111} \\
    log\text{ } |\dot{f}| &=&n\text{  }log\text{ } f+log\text{ }k 
    \label{eqnsss}
\end{eqnarray}
Equation \ref{diff_powlaws} forms lines in this space, where different braking indices correspond to lines with different slopes. If we consider a transformation $f \rightarrow$ $f+ \delta f$ and $k\rightarrow$ $k+dk$, we can find $dk$ such that the spindown remains constant when moving one frequency bin $\delta f=1/T_{FFT}$. After solving equations \ref{eqn111}-\ref{eqnsss} simultaneously, we find:

\begin{equation}
    dk=k\left( \left(1+\frac{\delta f}{f_{0,max}}\right)^{-n}-1\right)
    \label{dkeqn}
\end{equation}
Since $\delta f<< f_{0,max}$, we can Taylor expand to better understand the behavior of this grid:

\begin{equation}
        dk\approx -nk\frac{\delta f}{f_{0,max}}
\end{equation}
We can see that the grid is a function of the braking index and the value of $k$, so this grid is not uniform and changes for every Hough we do.

The minimum and maximum values for $k$ are related to the maximum and minimum spindowns we wish to analyze:

\begin{eqnarray}
k_{min}=\frac{\dot{f}_{0,min}}{f_{max}^n} \\
k_{max}=\frac{\dot{f}_{0,max}}{f_{min}^n}    
\end{eqnarray}

Combining these equations, we construct a grid on $k$ for each braking index, $T_{FFT}$ and source frequency.

To understand some of the properties of the grid on $k$, we plot in figure \ref{minmaxdk} how the range in $k$ and the step size $dk$ changes as a function of braking index:

\begin{figure}[ht!]
    \centering
    \includegraphics[width=80mm]{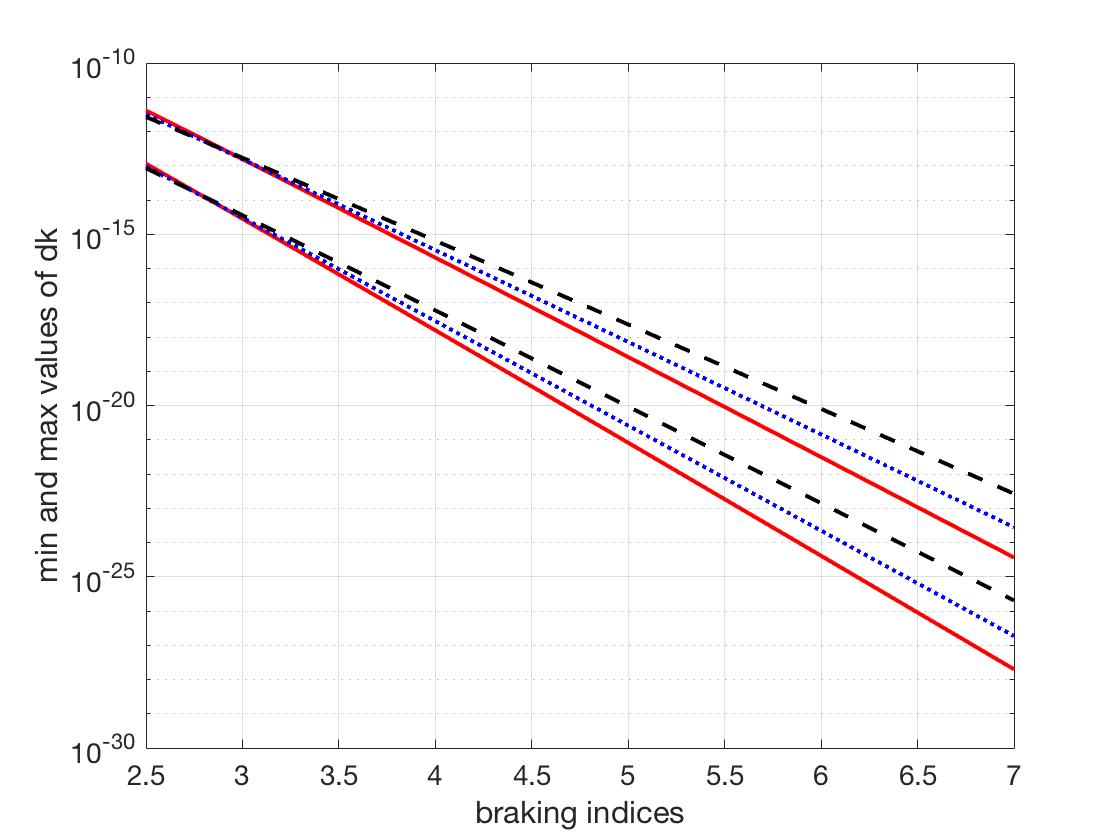}
    \caption{The ranges of the step $dk$ for different $T_{FFT}$. Red, blue, and black correspond to $T_{FFT}=2,4,8$ s (lowest, middle, highest curves) for frequency bands of [1000 2400] Hz, [600 1400] Hz, and [350 800] Hz respectively. $T_{obs}=5000 $ s and $\Delta t=1\times 10^6$ s.}
    \label{minmaxdk}
\end{figure}

\begin{figure}[ht!]
    \centering
    \includegraphics[width=80mm]{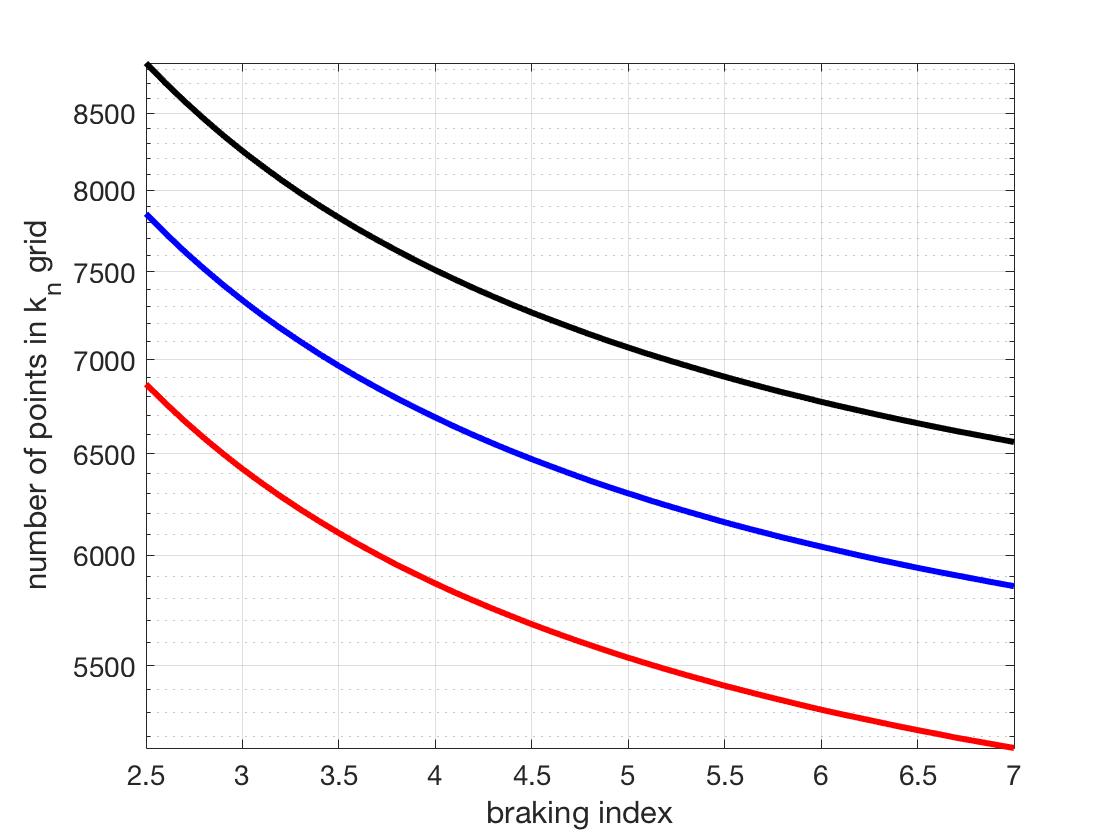}
    \caption{ The number of points in the grid on $k$ as a function of braking index. Less points are required to capture the variation in spindown at higher braking indices. Red (solid), blue (dots), and black (dashes) correspond to $T_{FFT}=2,4,8$ s for frequency bands of [1000 2400] Hz, [600 1400] Hz, and [350 800] Hz respectively. }
    \label{fig:k_pts}
\end{figure}

It is clear that $dk$ decreases with braking index, since the spindowns are smaller. However, the range $(k_{max}-k_{min})/dk$ decreases with increasing braking index, which means that we should expect fewer points in the grid on $k$ at higher frequencies. This is shown in figure \ref{fig:k_pts}.

\subsection{Grid on $x_0$}

The grid in $x_0$ is obtained by taking the derivative of Eq. \eqref{transform}:

\begin{equation}
dx_0=(n-1)\frac{\delta f}{f^{n}}
\end{equation}
In principle the grid on $x_0$ can change because $f$ varies; however this slows down the generalized FH greatly, especially at high frequencies. So we take the smallest step possible in $dx_0$, corresponding to the maximum frequency $f_{max}$ analyzed. This simply increases the size of each Hough map, but does not add computational time to the analysis, since all frequency bins are filled simultaneously at each spindown and time. 

\subsection{Splicing the parameter space}

We could explore the entire parameter space with a fine enough grid in $k$ and $n$; however this is not computationally practical nor is it necessarily the most sensitive. The signals we are searching for can have very high initial frequencies and enormous spindowns compared to continuous wave searches, which means that both are rapidly changing within hours, of $O(10^2)$ Hz and $O(10^{-1})$ Hz/s. The sensitivity $S$ of a semi-coherent search--a search in which one must break up the observation time into many chunks and combine their information-- is related to the duration of the signal we are looking for $T_{obs}$, the $T_{FFT}$ used, and the noise distribution of the detector:

\begin{equation}
    S\propto \frac{(T_{FFT}T_{obs})^{1/4}}{\sqrt{S_n}}f^2
    \label{sens}
\end{equation}
where $S_n$ is the noise power spectral density of the detector at a given frequency. Based on Eq. \eqref{sens}, we have found that we can improve the sensitivity by analyzing later times in a targeted search for signals of longer durations, which allows us to increase $T_{FFT}$. This technique has a few benefits: (1) the signal immediately after a merger is probably very complicated and not well-explained by models, (2) the signal's frequency will have decayed into a frequency band to which the detectors are more sensitive, and (3) we can probe higher initial frequencies and spindowns of the source that would otherwise lie outside of a good frequency band in the detectors. Depending on the total observation time and the frequency/spindown analyzed, we calculate how much time to cut from the analysis and subsequently what $T_{FFT}$ we can use to maximize sensitivity.

\begin{figure}[ht!]
    \centering
    \includegraphics[width=90mm]{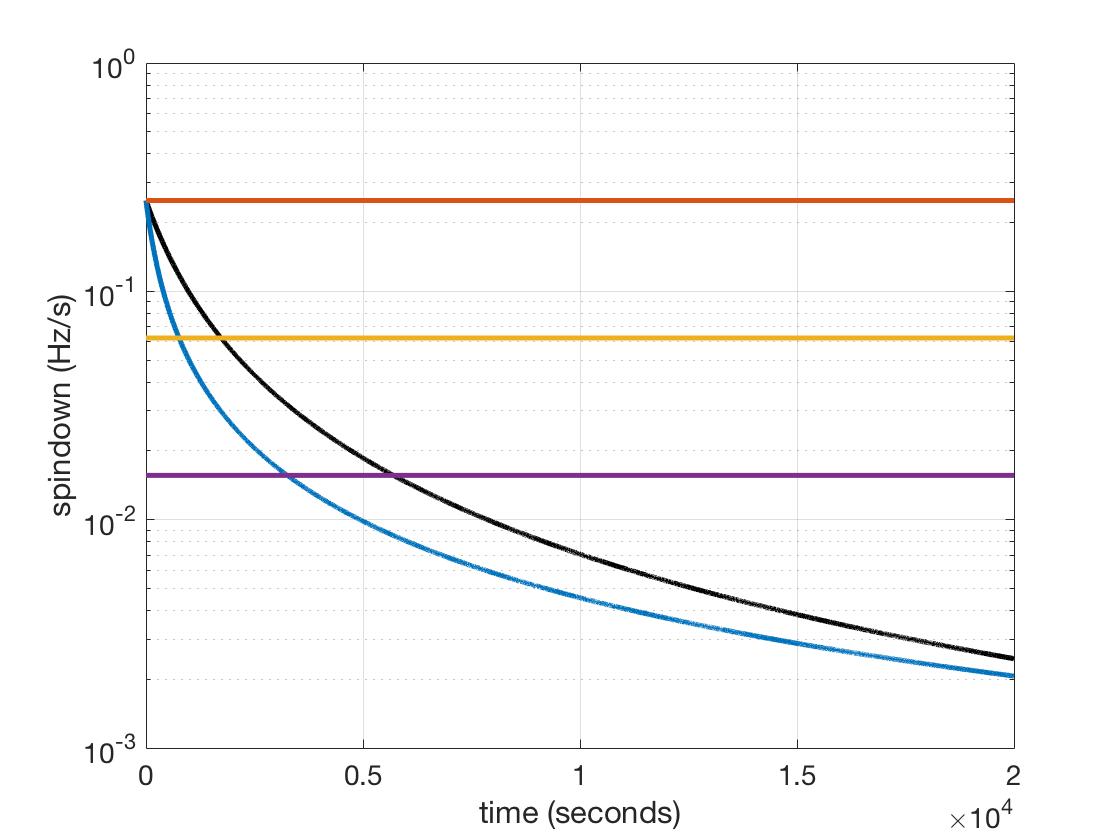}
    \caption{The spindown of two signals with $f_0=500$ Hz with different braking indices (black/top: $n=2.5$; red/bottom: $n=7)$ as a function of time. The horizontal lines correspond to spindowns $\dot{f}=1/T_{FFT}^2$, $\dot{f}=1/(2T_{FFT})^2$,  and $\dot{f}=1/(4T_{FFT})^2$. When a signal reaches a small enough spindown (intersects one of the horizontal lines), we can then analyze it for a longer duration with a higher $T_{FFT}$. Within each rectangle, the times and frequencies that will be analyzed with a given $T_{FFT}$, which will be different depending on $n,f_0$ and $\dot{f_0}$. We explore all signals whose paths lie between the curves corresponding to $n\geq2.5$ and $n\leq7$.}
    \label{change_tfft}
\end{figure}

However, at $T_{FFT}>=8$ s, the grid on $k$ becomes too large (see figure \ref{fig:k_pts}), and so now it is not practical for us to use longer $T_{FFT}$ given the current computational constraints. In order to perform a search, we decide to analyze different time and frequency bands depending on the spindown. A given $T_{FFT}$ implies that there is a maximum spindown such that the signal will remain within one frequency bin for the duration of $T_{FFT}$:


\begin{equation}
    \dot{f}=\frac{1}{T_{FFT}^2}
    \label{dot_tfft}
\end{equation}
Based on equations \ref{diff_powlaws} and \ref{powlaws}, there will be a time when the spindown becomes a factor of 4 smaller. After this time, we can safely increase $T_{FFT}$ by a factor of 2 while  still confining the power due to a signal to one (smaller) frequency bin. The time at which this occurs, and the frequency to which this corresponds, define the time and frequency bands to analyze for a given $T_{FFT}$, for a given braking index. We then perform the generalized FH in this frequency band, and continue to analyze the data with a larger $T_{FFT}$, shown in figure \ref{change_tfft}.

\begin{figure}[ht!]
    \centering
    \includegraphics[width=90mm]{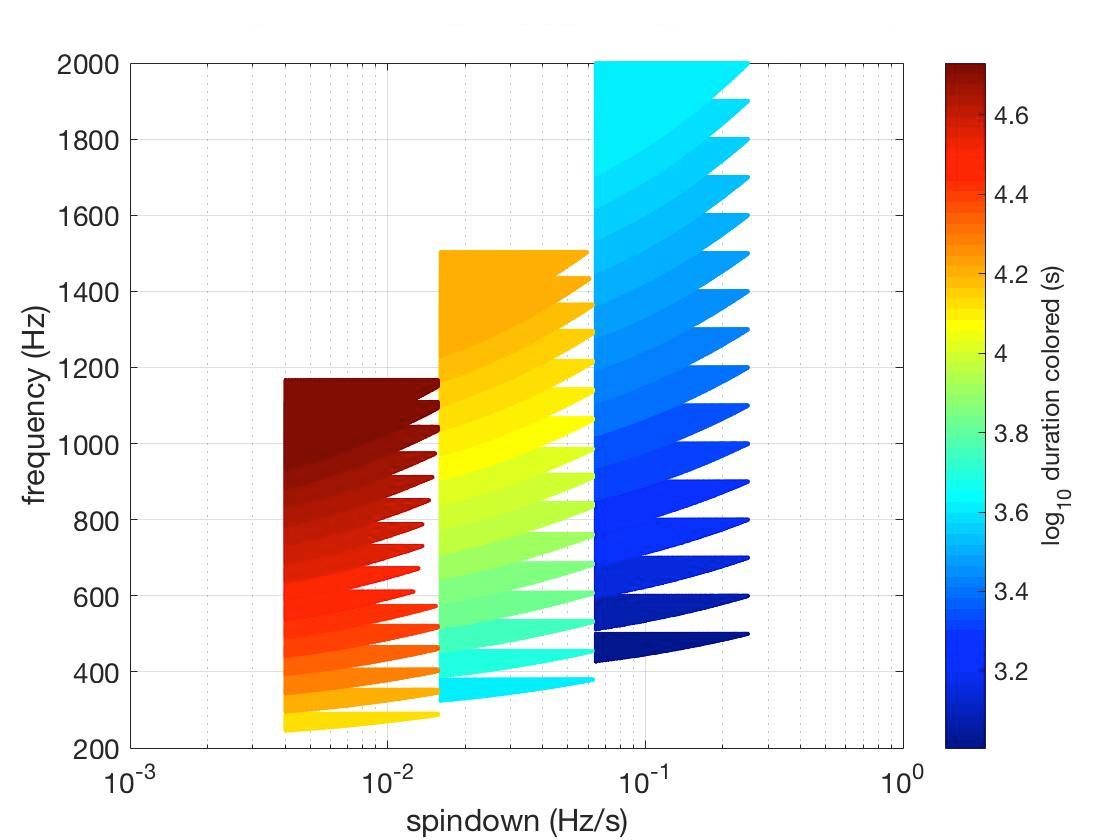}
    \caption{The frequency/spindown space that is explored by analyzing different portions of frequency and time with different $T_{FFT}$ for $n=5$. The durations $T_{obs}$ of the analysis are colored, and the three separate blocks correspond to $T_{FFT}=2,4,8$ s moving from right to left.}
    \label{parmspace}
\end{figure}

\begin{figure}[ht!]
    \centering
    \includegraphics[width=90mm]{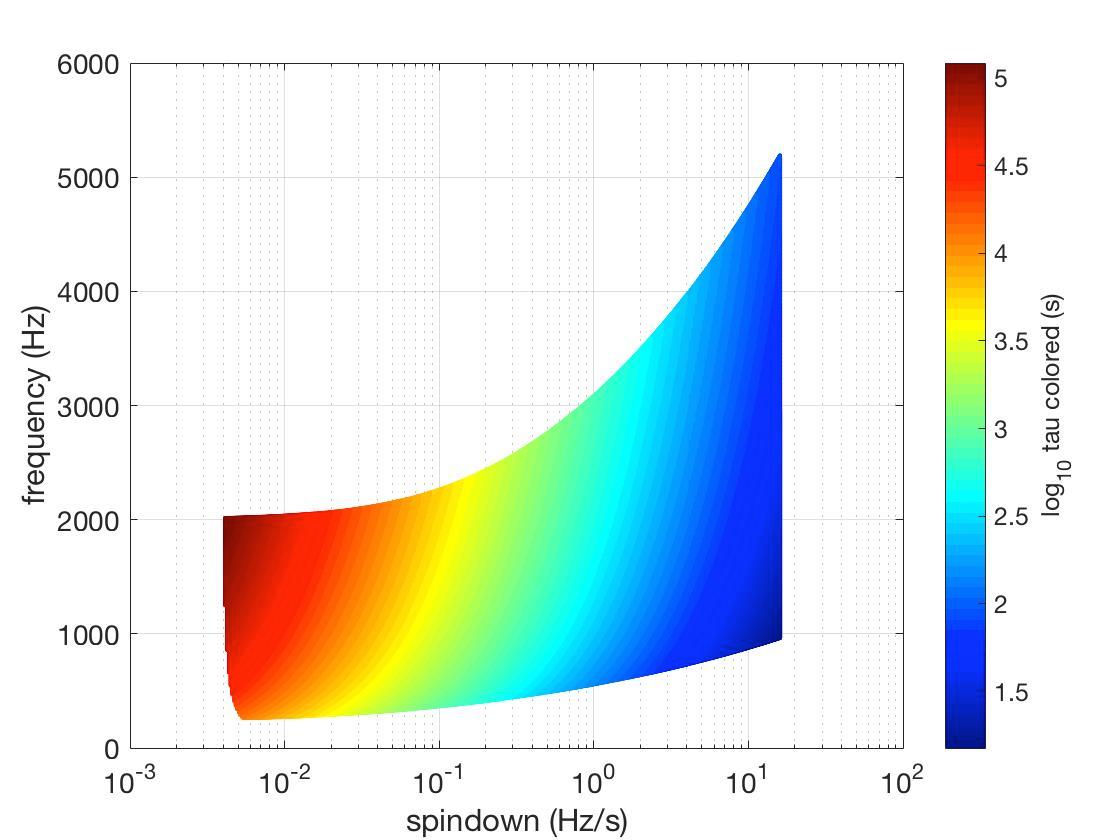}
    \caption{The source frequencies and spindowns that we are most sensitive to in our search scheme (figure \ref{change_tfft}), with a choice to start analyzing the data one hour after a merger, for $n=5$. The spindown timescale $\tau$ is colored. This plot includes signals that lie in the ``holes'' of figure \ref{parmspace}.}
    \label{sourceparmspace}
\end{figure}

For a given braking index $n=5$, we plot the portion of the $(f,\dot{f})$ space that is explored (see figure \ref{parmspace}). We are still sensitive to signals whose $f_0,/\dot{f_0}$ are in the ``holes'', but with a reduced signal-to-noise ratio, because the signals would have spun out of the frequency band that we are analyzing with the FH. If, however, we start the search a certain amount of time after a supernova explosion, or neutron star merger, we are actually probing sources with higher frequencies and spindowns. The initial parameters that we are most sensitive to are shown in figure \ref{sourceparmspace}, where the spindown timescale $\tau$ is defined as: 

\begin{equation}
    \tau =\frac{1}{k f_0^{n-1}(n-1)}=\frac{f_0}{\dot{f_0}(n-1)}
\end{equation}

To determine these initial parameters, we try many different initial frequency/spindown combinations to determine which parameters will fall within one frequency bin of the frequencies given in \ref{parmspace} after allowing this simulated signal to decay for a certain amount of time.


It is possible that a gravitational wave signal could be present in all portions of the space, just with a lower frequency/spindown at a later time. We can connect these portions of the parameter space by allowing candidates in the first portion of the parameter space (with $T_{FFT}=2$ s) to spindown and see if their frequencies, braking indices and $k$ values are within a few bins ($\sim 3$) of any of the candidates in the second portion of the parameter space. Essentially, we are doing coincidences between each portion of the parameter space we are analyzing. Based on our simple model, we expect that $k$ should not vary, nor should $n$, so we can use these facts to exclude certain candidates.

\section{\label{sec:level88}Sensitivity for long transient searches}
We derive an analytic expression for the sensitivity of a semi-coherent search for long transient periodic signals, such as those emitted by a newborn magnetar, assuming the analysis is done with the generalized FH transform. The expression is in fact a generalization of the sensitivity computed for standard continuous wave signals and given by Eq. (67) of \cite{fh_paper}.
We assume the signal is periodic but with a varying frequency and amplitude. A semi-coherent analysis is based on the condition that in each data segment, of length $T_{FFT}$, the signal frequency and amplitude are approximately constant. In particular, the frequency does not shift more than the frequency bin width $\delta f=1/T_{FFT}$. In order to compute an average sensitivity we first introduce the quantity
\begin{equation}
\Lambda=\frac{T_{FFT}}{2T_{obs}}\sum_{i=1}^{i=N} \lambda(f(t_i))
\label{Lambda}
\end{equation}
where $\lambda(f(t_i))$ is:
\begin{equation}
\lambda_i= \lambda(f(t_i))=\frac{4|\tilde{h}(f(t_i))|^2}{T_{FFT}S_n(f(t_i))}
\label{lambdati}
\end{equation}
where $\tilde{h}(f(t_i))$ is the Fourier transform of the gravitational wave signal, $N=T_{obs}/T_{FFT}$ is the number of FFTs used, and $S_n(f(t_i))$ is the detector noise power spectrum. The quantity $\Lambda$ is an average of the $\lambda_i$ over the observation window and satisfies the condition that if 
$\lambda_i=\mathrm{const}$ then $\Lambda=\mathrm{const}$, i.e. we are back to the standard situation of CW signals in which the frequency and the amplitude do not significantly change over the observation time.
As the signal amplitude $h_0$ varies with frequency as $f^2$ we write 
\begin{equation}
h_0(t_i)=\mathcal{A}f^2(t_i)=\mathcal{A}\mathcal{F}_i
\end{equation}
where:
\begin{equation}
    \mathcal{A}=\frac{4\pi^2GI_{zz}}{c^4}\frac{\epsilon}{d}
\end{equation}, 
This $\mathcal{A}$ which depends only the star ellipticity $\epsilon$, moment of inertia and distance $d$. 
We can then write
\begin{equation}
\Lambda \approx 4\frac{T_{FFT}}{2T_{obs}}\mathcal{A}^2\sum_i \frac{\mathcal{F}^2_i}{S_n(f_i)}\frac{2.4308}{25\pi}T_{FFT}
\label{Lambda2}
\end{equation}

Following the same procedure given in \cite{fh_paper}, we obtain:

\begin{widetext}
\begin{equation}
    \mathcal{A}_{min}=\frac{4.02}{N^{1/4}\theta_{thr}^{1/2}}\sqrt{\frac{N}{T_{FFT}}}\left(\sum_i \frac{\mathcal{F}^2_i}{S_n(f_i)}\right)^{-1/2}\left(\frac{p_0(1-p_0)}{p^2_1}\right)^{1/4}\sqrt{\left(CR_{thr}-\sqrt{2}\erfc^{-1}(2\Gamma)\right)}
    \label{h00min}
\end{equation}
\end{widetext}
where  $\theta_{thr}$ is the threshold for peak selection selection in the whitened spectra, $S_n$ is the noise spectral density of the detector, $p_0$ is the probability of selecting a peak above the threshold $\theta_{thr}$ if the data contains only noise , $p_1$ = $e^{−\theta_{thr}}$ − 2$e^{−2\theta_{thr}}$ + $e^{−3\theta_{thr}}$ , $CR_{thr}$ is the threshold we use to select candidates in the final FH map, and $\Gamma$ is the chosen confidence level.

The minimum detectable strain at a given confidence level can be obtained from Eq. \eqref{h00min} using a suitable ``frequency'' (indeed $h_{0,min}=\mathcal{A}_{min}\cdot \mathrm{frequency}^2$). We use the initial frequency $f_0$. 
The sensitivity depends on the signal evolution through the ratio $\sum_i \frac{\mathcal{F}^2_i}{S_n(f_i)}$. To calculate the sensitivity we must fix $n$, $f_0$, $\dot{f}_0$ and $T_{obs}$. That is, for a given detector and given search parameters ($T_{FFT},~\theta_{thr},~CR_{thr},~\Gamma$ etc.) and for each specific signal model, we will have a different sensitivity. 
If $f_i=const$, which is a good approximation for a standard CW signal case, then $\mathcal{F}_i=f_0^2$ and as a consequence $\sum_i \frac{\mathcal{F}^2_i}{S_n(f_i)}=\frac{Nf_0^4}{S_n(f)}$. Inserting this expression in Eq. \eqref{h00min}, we recover the standard sensitivity expression in \cite{fh_paper}.
If the emission of gravitational waves is due to r-modes, the signal amplitude scales with $f^3$. In this case $\mathcal{F}_i=f^3(t_i)$ and $\mathcal{A}=1.11\cdot 10^{-9}I\alpha$. 
The distance reach of the search is easily obtained inverting Eq. \eqref{h00min}:
\begin{widetext}
\begin{equation}
d_{max}=5.72\cdot 10^{-9}I_{38}\epsilon_{-3}\frac{T_{FFT}}{\sqrt{T_{obs}}}\left(\sum_i \frac{\mathcal{F}^2_i}{S_n(f_i)}\right)^{1/2}\left(\frac{p_0(1-p_0)}{Np^2_1}\right)^{-1/4}\sqrt{\frac{\theta_{thr}}{\left(CR_{thr}-\sqrt{2}\erfc^{-1}(2\Gamma)\right)}}
\label{dmax}
\end{equation}
\end{widetext}
with $I_{38}$ the star moment of inertia (with respect to the rotation axis) in units of $10^{38}kg\cdot m^2$ and $\epsilon_{-3}$ the star ellipticity in units of $10^{-3}$. For r-mode emission Eq. \eqref{dmax} must be modified replacing the numerical coefficient by $1.11\cdot 10^{-12}$ and it should be noted that $\epsilon $ is in fact the mode amplitude $\alpha$. 

\begin{figure}[ht!]
    \centering
    \includegraphics[width=90mm]{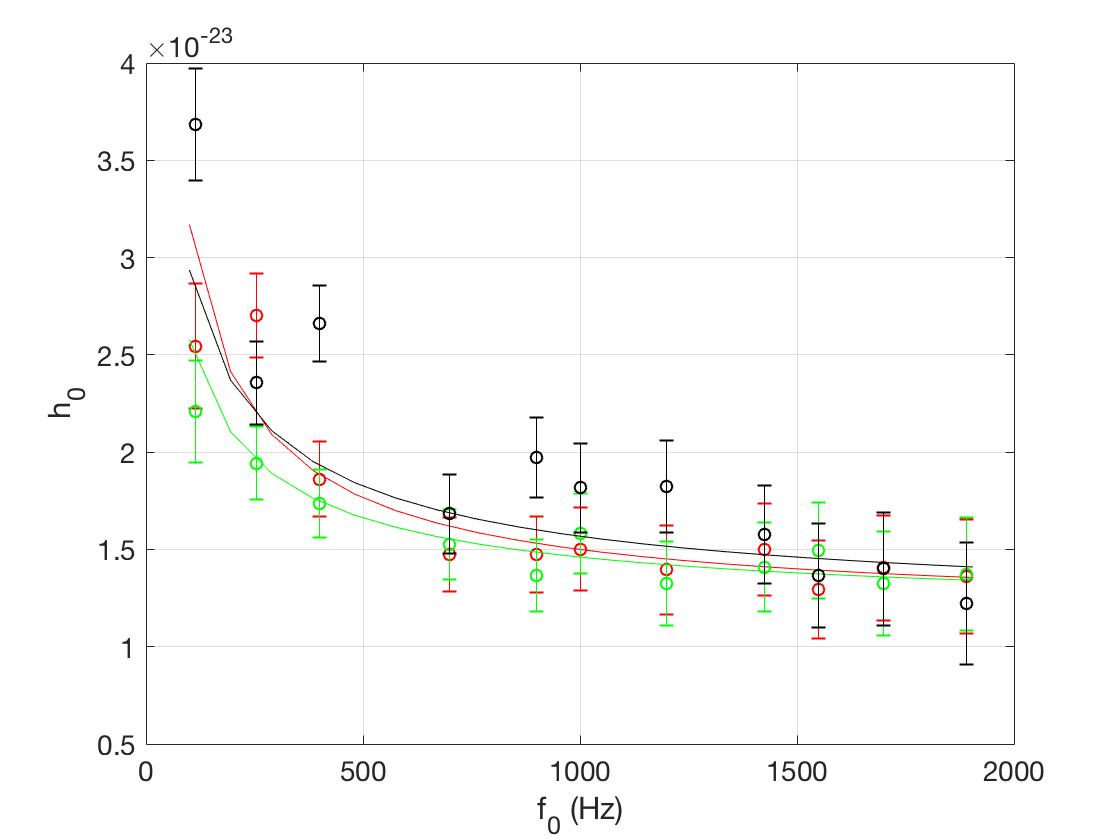}
    \caption{Minimum detectable amplitude $h_{0,min}$ as a function of initial frequency $f_0$ for braking indices $n=3,5,7$ (red, green and black / bottom, middle, and top). The continuous curves are the theoretical sensitivity estimates; the curves marked with circles are the sensitivity estimates obtained from injections in white noise at a level consistent with real O2 Livingston data about 1 hour after GW170817 ($S_n=7.94\times 10^{-24}$ $\frac{1}{\sqrt{Hz}}$); parameters: $T_{obs}=5000$ s, $T_{FFT}=4$ s, $\theta_{thr}=2.5$, $\dot{f_0}=1/T_{FFT}^2$, $\Gamma=0.9$. To compute the theoretical sensitivity estimates, we use a varying $CR_{thr}$ reflective of the average critical ratio we recover from injections at $90\%$ confidence.}
    \label{h0mincompar}
\end{figure}

\begin{figure}[ht!]
    \centering
    \includegraphics[width=90mm]{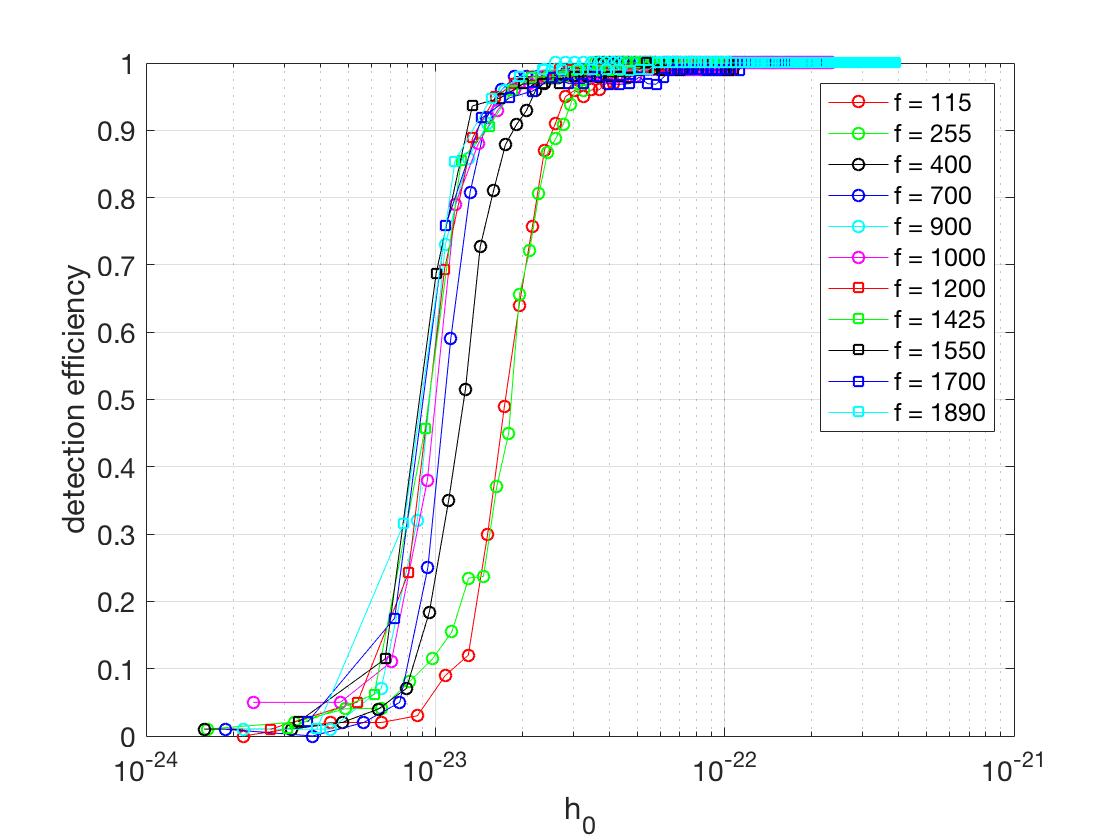}
    \caption{Detection efficiency plotted as a function of $h_0$ for $n=5$ for $T_{obs}=5000$ s and $T_{FFT}=4$ s. Different colors correspond to different initial frequencies $f_0$. The curves seem to follow the sigmoid distribution, which is as expected. We theoretically estimate the false alarm probability to be order of $0.01\%$ using Eq. 61 of \cite{fh_paper}; however the grids on $x_0$ and $k$ change in each Hough map, so the false alarm probability also changes. Moving from left to right tends to correspond to decreasing $f_0$.}
    \label{effvsampcurve}
\end{figure}

We empirically estimate the minimum amplitude detectable at $\Gamma=90\%$ confidence at some initial frequencies for fixed braking indices of $n=3,5,7$. We use 100 injections in Gaussian noise between $f_0$ and $f_0+10$ Hz and define a detection as when a candidate selected in a Hough map is within a distance of 3 $x_0/k$ bins from the injection. In a real search, there is a coincidence step between candidates selected in different detectors, but for this empirical sensitivity estimation, we only do coincidences between a candidate found in one detector and the injection. The error on each measurement is  $10\%$ of the theoretical $h_0$ for the corresponding $f_0$. The minimum detectable amplitudes are shown in figure \ref{h0mincompar} where the theoretical sensitivity curves for the same parameters are also plotted. The plot demonstrates a good agreement between experimental and theoretical results. Additionally, we show plots of detection efficiency as a function of amplitude for a few initial frequencies in figure \ref{effvsampcurve}. The false alarm probability is $\sim0.01\%$, but in general changes in each Hough map.

\begin{figure}[ht!]
    \centering
    \includegraphics[width=90mm]{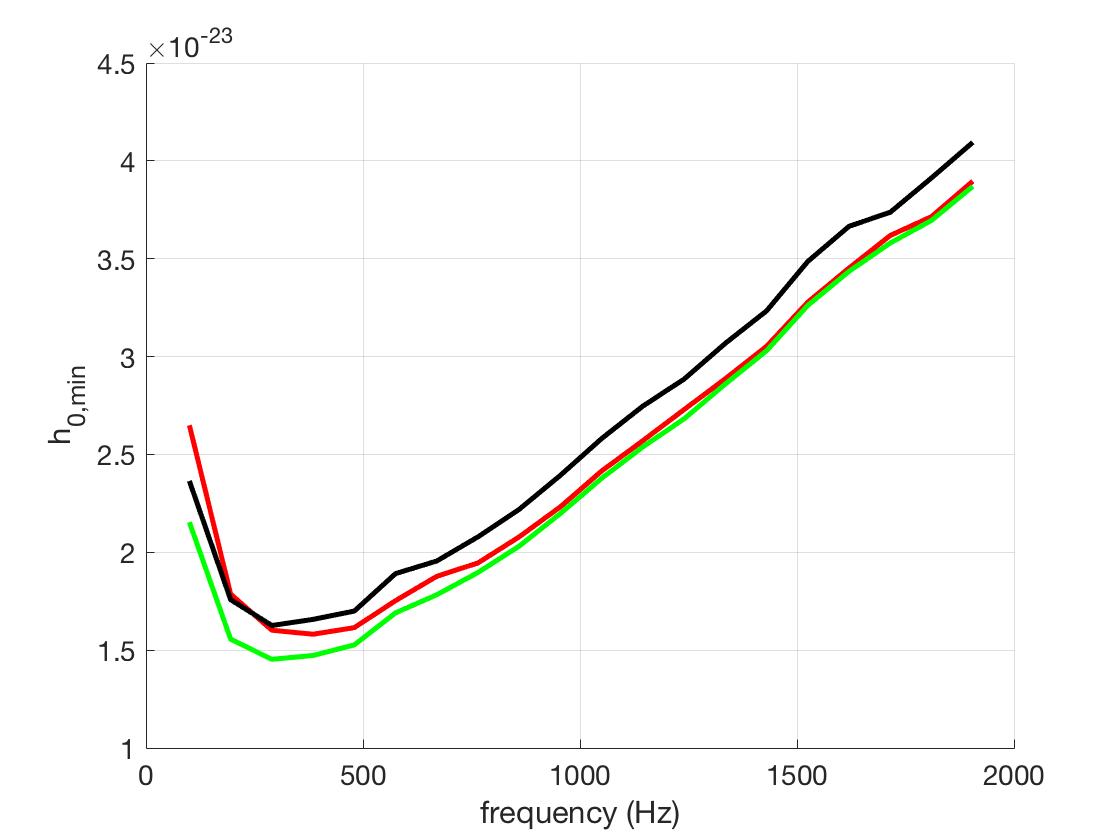}
    \caption{Theoretical $h_{0,min}$ for different initial frequencies $f_0$ for $n$=3, 5, 7 (red, green, black) where the noise distribution has been weighted with the O2 Livingston sensitivity curve for $\dot{f_0}=1/16$ Hz/s. 
    }
    \label{fig:sensi1}
\end{figure}

In this case the typical distance reach is of the order of 0.5-1 Mpc, obtained from the theoretical estimates shown in figure \ref{fig:sensi1} for O2 Livingston data, assuming a moment of inertia and star ellipticity.  The moment of inertia could be larger than the canonical $10^{38}~kg \cdot m^2$ for a neutron star born after a coalescence. The other parameter which may have some impact is the threshold for candidate selection $CR_{thr}$, which depends on how many candidate follow-ups we can afford. The distance reach is typically smaller for shorter $T_{obs}$. 

If we are able, thanks to observations in the EM band or to robust theoretical models, to restrict the possible range of the search parameters, $f_0,k,n$, we can in principle make a deeper search. For instance, by making a first rough search over a limited range of braking indexes we could make a second step searching over the braking index {\it residuals} (with respect to the initial rough grid) and this would allow us to increase $T_{FFT}$ and then gain in sensitivity.

\section{\label{sec:level11}Conclusions}

In this paper we have described a generalization of the FH transform, a method originally used in hierarchical continuous wave searches, to search for gravitational waves lasting $O(hours-days)$ originating from young, isolated neutron stars. We have shown that our method can be used to identify signals that have a frequency evolution in time that follows a power-law behavior. The sensitivity of our search has been computed theoretically and empirically, and the parameter space explored in a potential search has been discussed. We have also described how we perform a real search, and the way in which we follow up candidates.

Because the spindown of the neutron stars we search for is so high, we cannot use long $T_{FFT}$, nor can we be sure how long the signal will last in the detector band. Therefore we can only see $\sim 0.5-1$ Mpc away from us at current detector sensitivity. However, the Einstein Telescope is expected to be built and come online in the next decade, and with this instrument, a factor of $\sim 20$ improvement in sensitivity is expected \cite{et}. This means that we may have been able to see a remnant a source such as GW170817. Additionally, the development and implementation of new technology such as GPUs could help us expand our parameter space while reducing our computational cost.

We assume that the braking index is constant in time, but it is possible that different physical mechanisms are dominant at different times after the birth of a neutron star. Our method is able to handle this only if the variation of the braking index is small enough during the duration analyzed such that the frequency remains within one frequency bin. However, larger variations would result in signals either being lost completely or being recovered at a lower signal-to-noise ratio. To combat this problem, we plan to implement a machine-learning based method that is model-independent, using neural networks and random forests. The machine learning algorithms can be trained to recognize different constant or time-varying power laws, and have been shown to have a lot of success in detecting binary black hole mergers and glitches \cite{gabbard,george}. Since the physics after the birth of a neutron star is largely uncertain, machine learning could be a useful tool to analyze this portion of the parameter space. 

\section*{Acknowledgements}

We would like to thank the LIGO-Virgo Continuous Wave group for many useful discussions and the referees for reading our work.

\bibliographystyle{unsrt}


\end{document}